\documentstyle[psfig,eqsecnum,floats,preprint,aps]{revtex}
\newcommand{\PSbox}[3]{\mbox{\rule{0in}{#3}\includegraphics{#1}\hspace{#2}}}

\tighten

\begin{document}

\title{Quantum effects on winding configurations\\
in $SU(2)$-Higgs theory}

\author{Arthur Lue\footnote[1]{ithron@mit.edu}}

\smallskip

\address{Center for Theoretical Physics\\
Laboratory for Nuclear Science and\\
Department of Physics\\
Massachusetts Institute of Technology\\
Cambridge, MA\ \ 02139}

\maketitle

\vfill

\begin{abstract}

We examine the quantum corrections to the static energy for Higgs
winding configurations in order to ascertain whether such corrections
may stabilize solitons in the standard model.  We evaluate the
effective action for winding configurations in Weinberg-Salam theory
without $U(1)$-gauge fields or fermions.  For a configuration whose
size, $a \ll m^{-1}$ where $m = \max\{m_W,m_H\}$, $m_W$ is the W-mass,
and $m_H$ is the Higgs mass, the static energy goes like $g^{-2}m_W^2a
[1+b_0g^2\ln(1/ma)]^{c_0/b_0}$ in the semiclassical limit.  Here $g$
is the $SU(2)$-gauge coupling constant and $b_0, c_0$ are positive
numbers determined by renormalization-group techniques.  We discuss
the limitations of this result for extremely small configurations and
conclude that quantum fluctuations do not stabilize winding
configurations where we have confidence in $SU(2)$-Higgs as a
renormalizable field theory.

\end{abstract}

\setcounter{page}{0}
\thispagestyle{empty}

\vfill

\noindent CTP\#2585 \hfill Submitted to {\it Physical Review} {\bf D}

\noindent hep-th/9611059 \hfill Typeset in REV\TeX \eject

\baselineskip 24pt plus 2pt minus 2pt

\section{Introduction}

The Higgs sector in the standard model is a linear sigma model.  Such
a theory exhibits configurations of nontrivial winding, though they
are not stable.  In the standard model, winding configurations shrink
to some small size and then unwind via a Higgs zero when allowed to
evolve by the Euler-Lagrange equations.  These winding configurations
can be stabilized if one introduces four-derivative Higgs
self-interaction terms which are not present in the standard model
\cite{GipTze,AmbRub}.  The motivation typically cited for introducing
such terms is that one may treat the Higgs sector of the Lagrangian as
an effective field theory of some more fundamental theory which only
manifests itself explicitly at some high energy scale.  The stabilized
configurations have phenomenological consequences in electroweak
processes and provide an arena for testing nonperturbative aspects of
field theory and the standard model.  The presence of gauge fields
leads to the instability of these solitons \cite{AmbRub}, and their
decay, whether induced or by quantum tunnelling, is associated with
fermion number violation \cite{AmbRub,FGLR}.  In turn, such a
mechanism for fermion number violation may have an effect on the early
universe and electroweak baryogenesis \cite{baryo}.

Unfortunately, the procedure just described for stabilization is
inconsistent.  First, treating the Higgs sector as an effective theory
requires the inclusion of all higher-derivative terms of a given
dimension rather than just a few.  Moreover, using an effective field
theory to stabilize solitons creates difficulties.  Effective field
theories are equivalent to derivative expansions.  Solitons that are
stabilized by introducing the leading-order terms in a derivative
expansion imply that the following orders in the expansion will
contribute equally as the leading orders.  Truncating the derivative
expansion, then, is not legitimate.

We will take a different approach; we wish to see whether just the
quantum fluctuations of a {\em renormalizable} $SU(2)$-Higgs theory
can stabilize winding configurations.  We will take the Higgs sector
to be that found in the standard model.  Consider just the mode
associated with the size of a given winding configuration.  That mode
may be parametrized by the quantity $a$, the spatial size of the
object (which must be positive).  The potential for that degree of
freedom goes like $a+a^3$ and energetically favors configurations of
zero size.  Classically, the degree of freedom will evolve towards $a
= 0$ while losing its energy to radiative modes.  Now let us quantize
this mode.  Even though the system would favor being at $a=0$, that
would also imply the momentum conjugate to that degree of freedom
would also be zero.  Heisenberg uncertainty would puff out the
expectation value of the size degree of freedom.  Note that the size
of the soliton will be proportional to Planck's constant to some
power.  The mechanism described is analagous to that which stabilizes
the atom, which is classically unstable.  Of course, there are other
modes that couple into our situation, making the analysis more
complex.  Work has been done \cite{q-sol} that investigates quantizing
breathing modes in sigma models such as the Higgs sector of the
standard model.

In this paper, we identify the quantum effects on the energy of static
winding configurations by evaluating the effective action.  If quantum
effects stabilize solitons, that effect should be reflected by some
extremum in the effective action.  If we take the weak gauge-coupling
limit, $g^2 \rightarrow 0$, we shall see that an analytic expression
is available for the effective action that is a controlled
approximation in the regime where we have confidence in $SU(2)$-Higgs
theory as a renormalizable field theory.  The weak coupling limit is
equivalent to the semiclassical limit when fields are scaled properly.
When Planck's constant is small, we need only focus on small field
configurations.  This observation follows from the scenario described
above for stabilizing winding configurations by quantum fluctuations:
the size of a stabilized object will be proportional to Planck's
constant, and therefore, will be small if it exists.

Evaluating corrections to energies via the effective action is not a
new approach.  It has been a focus of investigation in effective meson
theories \cite{skyrm,small} and other areas \cite{phase-shift}.
Although, they elucidate important effects, such as bosonic and
fermionic back-reaction as well as valence fermion effects, these
studies treat the effective action to one-loop order only.  We will
see that there is a difficulty with simultaneously examining small
configurations and neglecting higher-loop effects.

We begin by spelling out the action and types of configurations under
consideration.  When configuration sizes are small, we can employ
renormalization-group techniques to ascertain the asymptotic behavior
of the one-particle irreducible green's functions which, in turn,
allows us to determine the leading-order size dependence of the
effective action.  We will find our results to be reliable until the
size of our configuration becomes extremely small, on the order of the
inverse momentum of the Landau pole associated with the Higgs sector.
Thus our results are valid when the size of our configuration is not
on the order of the cutoff necessary to avoid a trivial renormalized
Higgs self-coupling.  We conclude that the quantum corrections to the
energy are not sufficient to stabilize Higgs winding configurations
where we have confidence in $SU(2)$-Higgs as a renormalizable field
theory.

\section{The Effective Action}

Consider the Weinberg-Salam theory of electroweak interactions,
neglecting the $U(1)$-gauge fields and fermions.  Our field variables
form the set $\{A_\mu(x),\phi(x)\}$ where the gauge field $A_\mu(x) =
\sigma^aA_{\mu a}(x)/2$ is in the adjoint representation of $SU(2)$
($\{\sigma^a\}$ are the Pauli matrices), and the Higgs field $\phi(x)$
is in the fundamental representation of $SU(2)$.  The classical action
we consider is
\begin{eqnarray}
	S &=& S_0 + S_{gf} \nonumber \\
	S_0 &=& \int d^4x \left\{
		-\frac{1}{4} F^{\mu\nu}_aF^a_{\mu\nu} + \left({\cal
		D}^\mu\phi\right)^\dagger \left({\cal D}_\mu\phi\right) -
		\frac{g^2\alpha}{2}\left[(\phi+\phi_0)^\dagger(\phi+\phi_0)
		-\frac{m_W^2}{g^2}\right]^2\right\} \label{su2-action}  \\
	S_{gf} &=& -\frac{1}{2\xi}\int d^4x \left[
		\partial_\mu A^{\mu a} + ig^2\xi
		\left(\phi^\dagger\sigma^a\phi_0 -
		\phi^\dagger_0\sigma^a\phi\right)\right]^2 \nonumber
\end{eqnarray}
where we have chosen the $R_\xi$-gauge and
\begin{eqnarray}
	{\cal D}_\mu\phi &=& (\partial_\mu -
				\frac{ig}{2}\sigma^aA_{\mu a})\phi,
				\nonumber \\ F^a_{\mu\nu} &=&
				\partial_\mu A^a_\nu - \partial_\nu
				A^a_\mu + g\epsilon^{abc}A_{\mu
				b}A_{\nu c}.  \nonumber
\end{eqnarray}
We have shifted $\phi(x)$ by a vacuum configuration $\phi_0$, some
constant $SU(2)$-doublet satisfying the relationship
$\phi_0^\dagger\phi_0 = m_W^2/g^2$.  The parameter $m_W$ is the mass of
the gauge field and $\alpha = (m_H/m_W)^2$ where $m_H$ is the
physical Higgs mass.  The Feyman rules derived from (\ref{su2-action})
are familiar.  In this analysis, we restrict ourselves to the
semiclassical limit, which is equivalent to taking $g^2 \rightarrow 0$
while holding $m_W, m_H$ (and, thus, also $\alpha$) fixed.

We wish to determine the effects of quantum fluctuations on Higgs
winding configurations.  We will evaluate the effective action,
$\Gamma[A^a_\mu,\phi]$, where
\begin{eqnarray}
	A^a_\mu(x) &=& 0	\nonumber	\\
	\phi(x) &=& \left[U({\bf x}) - 1\right]\phi_0
\label{winding}
\end{eqnarray}
and where $U({\bf x}) \in SU(2)$ is a static configuration such that
$U({\bf x}) \rightarrow 1$ as $|{\bf x}| \rightarrow \infty$ with
characteristic size, $a$.  We require the field $U({\bf x})$ to be a
configuration of unit winding number
$$
	w[U] = \frac{1}{24\pi^2}\int d^3{\bf x}\ \epsilon^{ijk}
		{\rm Tr}\left[(U^\dagger\partial_iU)
		(U^\dagger\partial_jU)(U^\dagger\partial_kU)\right] = 1.
$$

From conventional perturbation theory, we know that
\begin{equation}
	\Gamma[\phi] = \sum_n \int\frac{d^4p_1}{(2\pi)^4}\cdots
			\frac{d^4p_n}{(2\pi)^4}(2\pi)^4
			\delta^4\left(\sum_ip_i\right)
			\Gamma^{(n)}(\{p_i\})\hat\phi(p_1)\cdots\hat\phi(p_n)
\label{effaction}
\end{equation}
where $ \Gamma^{(n)}(\{p_i\})$ are the one-particle irreducible
green's functions with $n$ external $\phi$'s.  We define the field
$\hat\phi(p)$ as the fourier transform of $\phi(x)$.  For the
configuration (\ref{winding}), we get
$$
	\hat\phi(p) = 2\pi\delta(p_0)\int d^3{\bf x}\ 
		e^{-i{\bf p}\cdot{\bf x}}\phi({\bf x}).
$$
The static energy for the state whose expectation value of the operator
associated with the Higgs field is $\phi({\bf x})$ will be the quantity $E$ in
the expression
$$
	\Gamma[\phi] = -\int dt\ E.
$$
Normally, the effective action (\ref{effaction}) cannot be evaluated exactly.
However, because we are investigating the semiclassical limit, we are only
interested in a configuration whose size, $a$, is small.  We find that under
such a circumstance, we may use the Callan-Symanzik equation for our theory to
evaluate the leading-order size dependence of the one-particle irreducible
green's functions in (\ref{effaction}) and thus evaluate the quantum
corrections to the static energy.

\section{Asymptotic Behavior}

In this section we identify the scaling dependence of the one-particle
irreducible green's functions, $\Gamma^{(n)}(\{p_i\})$, when the
characteristic size of the Higgs winding configuration, $a$, is small
with respect to both mass scales of the theory.  Let $m =
\max\{m_W,m_H\}$.  We implement the condition of small, static
background configurations by requiring the field $\hat\phi(p)$ to have
support only for $p_0 = 0$ and $|{\bf p}| \gg m$ which implies $0 <
m^2 \ll -p^2$.  Under this circumstance, the asymptotic dependence of
$\Gamma^{(n)}$ will be determined by the Callan-Symanzik equation
\cite{CL}
\begin{equation}
	\left[\frac{\partial}{\partial s}
		- \beta_g(g^2,\alpha)\frac{\partial}{\partial g^2}
		- \beta_\alpha(g^2,\alpha)\frac{\partial}{\partial\alpha}
		+ n\gamma(g^2,\alpha) + (n-4)\right]
		\Gamma^{(n)}(\{e^s\tilde p_i\},g^2,\alpha,m) = 0.
\label{C-S}
\end{equation}
where $s = \ln(1/ma)$ and $-{\tilde p_i}^2 \sim m^2$.  Here $\beta_g,
\beta_\alpha$ are the beta functions for $g^2$ and $\alpha$,
respectively, that determine the running couplings.  The function
$\gamma(g^2,\alpha)$ is the anomalous dimension of the scalar field
$\phi(x)$.

The solution to (\ref{C-S}) may be determined by the method of characteristics.
We find that
\begin{eqnarray}
	\Gamma^{(n)}_{asymp} =
	(ma)^{n-4}\Gamma^{(n)}(\{\tilde p_i\},
	\bar{g}^2(s,g^2,\alpha),
	\bar{\alpha}(s,g^2,\alpha),m)\ \ \ \ \ \ \ \ \ \ \ \ \ \ \ \nonumber\\
	\ \ \ \ \ \ \ \ \ \ \times\exp\left[-n\int_0^sd\sigma\ 
	\gamma(\sigma,\bar{g}^2(\sigma,g^2,\alpha),
	\bar{\alpha}(\sigma,g^2,\alpha))\right]
\label{asymp}
\end{eqnarray}
where the running coupling constants $\bar g^2, \bar\alpha$ are defined such
that
$$
	\frac{d\bar g^2}{ds} = \beta_g(\bar g^2,\bar\alpha),\ \ \ \ \ \ 
	\frac{d\bar\alpha}{ds} = \beta_\alpha(\bar g^2,\bar\alpha)
$$
and $\bar g^2(s=0) = g^2, \bar\alpha(s=0) = \alpha$.  We will evaluate
(\ref{asymp}) using one-loop beta functions and scalar anomalous dimension.

The one-loop beta functions may be easily obtained from the literature.
For a single scalar Higgs doublet, we find \cite{GW}
\begin{eqnarray}
	\beta_g(g^2,\alpha) &=& -b_0g^4
\label{beta_g} \\
	\beta_\alpha(g^2,\alpha) &=& g^2(A\alpha^2+B\alpha+C)
\label{beta_alpha}
\end{eqnarray}
where $b_0 = 43/48\pi^2, A = 3/4\pi^2, B = 1/3\pi^2, C = 9/64\pi^2$.
\begin{figure}\begin{center}\PSbox{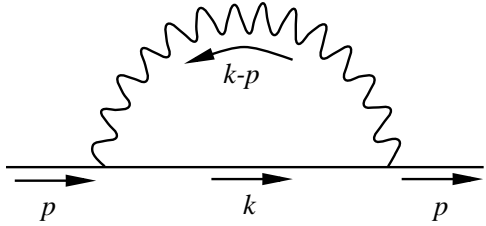}{2in}{1in}\end{center}
\caption{One-loop contribution to the scalar field anomalous dimension.}
\label{fig:figure1}
\end{figure}
The one-loop anomalous dimension may be evaluated using just the graph in
Figure \ref{fig:figure1}.
\begin{equation}
	\gamma(g^2,\alpha) = -\frac{1}{2}c_0g^2
\label{gamma}
\end{equation}
where $c_0 = 3[1+(\xi-1)/4]/16\pi^2$.  Because $\xi$ is positive or zero,
$c_0 \ge 9/64\pi^2$.  Inserting these expressions into (\ref{asymp}), we find
that
$$
	\Gamma^{(n)}_{asymp} = (ma)^{n-4}
	\Gamma^{(n)}(\{\tilde p_i\},\bar g^2, \bar\alpha)
	\left(1+b_0g^2s\right)^\frac{nc_0}{2b_0}.
$$
When $n>4$, the one-loop $n$-point diagrams give the leading terms for the
prefactor
\begin{equation}
	\Gamma^{(n)}(\{\tilde p_i\},g^2,\alpha) \sim m^{4-n}g^nF_n(\sqrt\alpha)
\label{1pi-n}
\end{equation}
where $F_n$ is a polynomial in $\sqrt\alpha$ of order $n$.  When $n \le 4$,
the renormalized $\Gamma^{(n)}$ is dominated by the classical contribution
\begin{equation}
	\Gamma^{(2)}(\tilde p,g^2,\alpha) \sim m^2,\ \ \ 
	\Gamma^{(3)}(\{\tilde p_i\},g^2,\alpha) \sim m_Wg\alpha,\ \ \ 
	\Gamma^{(4)}(\{\tilde p_i\},g^2,\alpha) \sim g^2\alpha.
\label{1pi-234}
\end{equation}
We are now prepared to insert these asymptotic formulas into
(\ref{effaction}).  So long as $\bar\alpha/\alpha$ is not too large,
the leading-order size dependence of the effective action comes from
the two-point one-particle irreducible greens function.  All other
terms are supressed by powers of $a$ and other factors.  The
leading-order contribution to the effective action from quantum
fluctuations yields
\begin{equation}
	\Gamma[\phi] = \int\frac{d^4p}{(2\pi)^4}\phi^\dagger(p)\phi(p)
		p^2\left[1+\frac{1}{2}b_0g^2\ln\left(\frac{-p^2}{m^2}\right)
		\right]^\frac{c_0}{b_0}
\label{2-point}
\end{equation}
implying a scale dependence
\begin{equation}
	\Gamma[\phi] \sim -\int dt\ \frac{m_W^2a}{g^2}
		\left[1+b_0g^2\ln\left(\frac{1}{ma}\right
		)\right]^\frac{c_0}{b_0}.
\label{energy}
\end{equation}
One can recover the classical result from (\ref{energy}) by setting
the $g^2$ inside the brackets to zero.  Note that when
$b_0g^2\ln(\frac{1}{ma}) \sim 1$, the quantum corrections to the
energy are as significant as the classical contribution.
Nevertheless, the static energy that corresponds to this effective
action is a monotonically increasing function of the size, $a$, such
that $E(a=0) = 0$\footnote{Note that the expressions (\ref{2-point})
and (\ref{energy}) are dependent on the gauge fixing parameter, $\xi$,
through the exponent $c_0/b_0 = 9[1+(\xi-1)/4]/43$.  The effective
action will generally depend on $\xi$ when the background
configuration is not an extremum of the effective action
\cite{e-action}.  The effective action need only be gauge-parameter
independent when $\delta\Gamma/\delta\phi = 0$, when the field
configuration is some solution to the quantum-corrected equations of
motion.  Our observation that the static energy has no extrema for
small-sized configurations is not altered by different choices of
$\xi$.}.  This would imply that Higgs winding configurations would
shrink to zero size and unwind via a Higgs zero, just as in the
classical scenario.

Let us take a closer look at our expression for the leading contribution to the
effective action (\ref{2-point}).  Expanding in powers of $g^2$ we get
$$
		\Gamma[\phi] = \int\frac{d^4p}{(2\pi)^4}\phi^\dagger(p)\phi(p)
		p^2 + \frac{c_0}{2}g^2\int\frac{d^4p}{(2\pi)^4}
		\phi^\dagger(p)\phi(p)p^2\ln\left(\frac{-p^2}{m^2}\right)
		+ \cdots
$$
The first term is the contribution from the classical action.  The next term is
the leading order contribution from one-loop one-particle irreducible graphs.
Note that it depends only on $c_0$.  In fact, the dominant one-loop
contribution to the effective action comes from the graph in Figure
\ref{fig:figure1}.  The scale dependence of the static energy will go like
\begin{equation}
	E = \frac{m_W^2a}{g^2}\left[{\cal A} + {\cal B}g^2\ln\frac{1}{ma}
		+ {\cal C}g^4\left(\ln\frac{1}{ma}\right)^2 + \cdots\right]
\label{expansion}
\end{equation}
where $\cal A, \cal B, \cal C$ are numbers.  Again the first term is
the classical energy, the second is the one-loop energy, and the rest
of the terms in the expansion (\ref{expansion}) correspond to
higher-loop energies order by order.  We can see by comparing
(\ref{energy}) with (\ref{expansion}) that loop contributions to the
effective action beyond one loop can only be neglected when
$b_0g^2\ln(1/ma) \ll 1$.  However, that is precisely the condition
where the one-loop contribution can be neglected relative to the
classical action.  Thus, drawing conclusions concerning solitons based
on one-loop results may be difficult.  When dealing with small
configurations, one still needs to include higher-loop contributions,
even in the semiclassical limit.

\section{Limitations}

We expect the expression (\ref{energy}) to be valid so long as we can
neglect high-loop contributions to the beta functions
(\ref{beta_g}--\ref{beta_alpha}), the anomalous dimension
(\ref{gamma}), and the renormalized greens functions
(\ref{1pi-n}--\ref{1pi-234}).  Higher loop contributions to these
quantitities will be negligible if the running couplings associated
with the gauge coupling, $\bar g^2$, and that associated with the
scalar self-coupling, $\bar g^2\bar\alpha$, are small.  The first
criterion is relatively easy to adhere to, since the gauge coupling is
asymptotically free and will run to zero with increasingly small sized
configurations, subject to $\bar g^2\bar\alpha$ being small. Recall
also, we require that $\bar\alpha/\alpha$ not be too large so that we
may neglect all but the two-point contribution to the effective
action.  All these criteria for the validity of (\ref{energy}) hinge
on how $\bar\alpha$ runs with smaller and smaller sized
configurations.  From (\ref{beta_alpha}) we can ascertain the
dependence of $\bar\alpha$ on the configuration size.  Inserting the
appropriate expression for $\bar g^2$, we find
\begin{equation}
	\bar\alpha(s,g^2,\alpha) =
	\frac{1}{2A}\left\{-B+D\tan\left[\tan^{-1}\frac{2A\alpha+B}{D}
			+ \frac{D}{2b_0}\ln(1+b_0g^2s)\right]\right\}
\end{equation}
where $D = \sqrt{4AC-B^2} > 0$.  Thus we see that $\bar\alpha$ has a
Landau singularity when the argument of the tangent is $\pi/2$.  Thus,
the expression (\ref{energy}) is certain to break down for a
configuration of some size approaching the inverse momentum associated
with this singularity.  However, if the argument of the tangent is not
very close to $\pi/2$, then $\bar\alpha$ will not be numerically very
different from the renormalized $\alpha$, implying that $\bar
g^2\bar\alpha$ is small if $g^2\alpha$ is small.  Thus, our constraint on
the size, $a$, will be that
$$
	\tan^{-1}\frac{2A\alpha+B}{D}
			+ \frac{D}{2b_0}\ln(1+b_0g^2s) < \frac{\pi}{2}.
$$
This implies that
\begin{figure}\begin{center}\PSbox{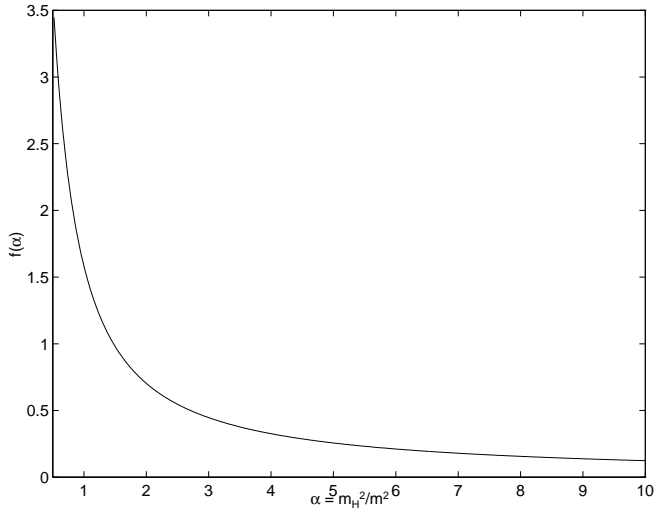 hoffset=-50 voffset=-100
hscale=50 vscale=50}
{3in}{2.5in}\end{center}
\caption{The function $f(\alpha)$ is plotted versus $\alpha =
{m_H}^2/m^2$.  The scale of the Landau pole may be extracted from this
curve using the identity $f(\alpha) = b_0g^2\ln(\Lambda_{Landau}/m)$.
The criterion for the validity of our expression for the quantum-corrected
energy is that $a > 1/\Lambda_{Landau}$ or, alternatively,
$a > m^{-1}\exp[-f(\alpha)/b_0g^2]$.}
\label{fig:figure2}
\end{figure}
$$
	b_0g^2s < e^{\frac{2b_0}{D}\left[\frac{\pi}{2}
			-\tan^{-1}\frac{2A\alpha+B}{D}\right]} - 1
			\equiv f(\alpha).
$$
The function, $f(\alpha)$ is plotted in Figure \ref{fig:figure2}.  Thus, our
expression (\ref{energy}) is valid for configurations whose size, $a$,
satisfies
\begin{equation}
	m^{-1}\ e^{-f(\alpha)/b_0g^2} \ll a \ll m^{-1}
\label{constraint}
\end{equation}
when $g^2, g^2\alpha \ll 1$.  Recall from our discussion at the end of the
last section that the quantum corrections to the energy will be as
significant as the classical contribution when $b_0g^2s \sim 1$.  Now
we can see from our condition $b_0g^2s < f(\alpha)$ and Figure
\ref{fig:figure2} that there exists some $\alpha_{max}$ such that if
$\alpha < \alpha_{max}$, the energy (\ref{energy}) will be both
reliable and significantly different than the classical result.  When
$\alpha > \alpha_{max}$ the quantum corrections are bound to be small
where our result is valid.

When the size of the configuration is near the inverse momentum
associated with the Landau pole, higher-loop contributions to the beta
and gamma functions become important and potentially change the
size-dependence of the energy.  However, if we take the presence of
this singularity as a cue that a finite momentum cutoff is necessary
to maintain a nontrivial Higgs self-coupling \cite{rg}, then we can
conclude that our expression for the static energy (\ref{energy}) is
valid so long as the size of the configuration is not so small as to
be on the order of a lattice size associated with a finite momentum
cutoff.  Thus, quantum fluctuations do not stabilize winding
configurations where we have confidence in $SU(2)$-Higgs theory as a
renormalizable field theory.

\section{Concluding Remarks}

We evaluate the quantum corrections to the static energy for a Higgs
winding configuration in the semiclassical limit.  We ascertain the
leading-order contributions to the static energy which is relevant to
soliton stabilization by treating the effective action for small-sized
configurations.  Moreover, we perform this calculation in such a way
that avoids the difficulties in neglecting higher-loop contributions
to the effective action.  By solving the Callan-Symanzik equation for
our theory, we find the leading-order size dependence of the energy
including quantum corrections.  We also use renormalization-group
techniques to determine the limitations on our expression.

We find that bosonic quantum fluctuations do not stabilize Higgs
winding configurations in a standard model type $SU(2)$-Higgs theory.
Classically, a configuration of size, $a$, would have static energy
which goes like $m_W^2a/g^2$.  These winding configurations will
shrink to zero size and eventually unwind via a Higgs zero.  Including
quantum fluctuations, when $a \ll m \equiv \max\{m_W,m_H\}$, the
corrected energy goes like
$g^{-2}m_W^2a[1+b_0g^2\ln(1/ma)]^{c_0/b_0}$, where $b_0, c_0$ are
positive numbers.  The corrected energy still goes to zero as the
size, $a$, goes to zero.  When the configuration shrinks to a size, $a
\sim m^{-1} \exp[-f({m_H}^2/m_W^2)/b_0g^2]$, our result becomes invalid
and we run into the Landau pole associated with the Higgs sector.
Taking this pole to be an indication that a finite momentum cutoff is
necessary for a nontrivial Higgs self-coupling, we find that bosonic
quantum fluctuations do not stabilize Higgs winding configurations
where we have confidence in $SU(2)$-Higgs theory as a renormalizable
field theory.

\acknowledgments

The author wishes to express gratitude for the help of E. Farhi in this work.
The author also wishes to acknowledge helpful conversations with J. Goldstone,
 K. Rajagopal, K. Huang, K. Johnson, L. Randall, M. Trodden, and T. Schaefer.
This work 
was supported by funds provided by the
U.S.~Department of Energy (D.O.E.)
under cooperative agreement \#DF-FC02-94ER40818.


\begin{references}

\bibitem{GipTze}J. M. Gipson and H. C. Tze, Nucl. Phys.
{\bf B183}, 524 (1981);\\ 
J. M. Gipson, {\it ibid} {\bf B231}, 365 (1984). 
\medskip

\bibitem{AmbRub}J. Ambjorn and V. A. Rubakov, Nucl. Phys. {\bf B256}, 
434 (1985);\\
V. A. Rubakov, Nucl. Phys. {\bf B256}, 509 (1985).
\medskip

\bibitem{FGLR}
V. A. Rubakov, B. E. Stern, P. G. Tinyakov, Phys. Lett. {\bf 160B}, 292 (1985);
\\
E. Farhi, J. Goldstone, A. Lue, and K. Rajagopal,
Phys. Rev. D {\bf 54}, 5336 (1996). 
\medskip

\bibitem{baryo}V. A. Kuzmin, V. A. Rubakov and M. E. Shaposhnikov,
Phys. Lett. {\bf 155B}, 36 (1985);\\
P. Arnold and L. McLerran, Phys. Rev. D {\bf 36}, 581 (1987).
For reviews, see N. Turok, in {\it Perspectives in Higgs Physics},
G. Kane, ed., World Scientific, 300 (1992); 
and A. Cohen,
D. Kaplan and A. Nelson, Ann. Rev. Nucl. Part. Phys. {\bf 43}, 27 (1993).
\medskip

\bibitem{q-sol}P. Jain, J. Schechter, and R. Sorkin, Phys. Rev.D {\bf 39}, 998
 (1989);\\
A. Kobayashi, H. Otsu and S. Sawada, Phys. Rev. D {\bf 42}, 1868 (1990);\\
N.M. Chepilko, K. Fujii and A.P. Kobushkin, Phys. Rev. D {\bf 43}, 2391 (1991).
\medskip

\bibitem{skyrm}
G. Ripka and S. Kahana, Phys. Lett. {\bf 155B}, 327 (1985);\\
I.J.R. Aitchison and C.M. Fraser, Phys. Rev. D {\bf 32}, 2190 (1985);\\
B. Moussallam, Phys. Rev. D {\bf 40}, 3430 (1989), hep-ph/9211229.
\medskip

\bibitem{small}
J. Baacke and A. Schenk, Z. Phys. {\bf C37}, 389 (1988);\\
J. A. Zuk, Int. J. Mod. Phys. {\bf A7}, 3549 (1992)
\medskip

\bibitem{phase-shift} C. L. Y. Lee, Phys. Rev. D {\bf 49}, 4101 (1994);
M. Bordag, J. Phys. A {\bf 28}, 755 (1995);\\
M. Bordag and K. Kirsten, Phys. Rev. D {\bf 53}, 5753 (1996).
\medskip

\bibitem{CL}
T.-P. Cheng and L.-F. Li, {\em Gauge theory of elementary particle physics},
Oxford (1984).
\medskip

\bibitem{GW}
D. J. Gross and F. Wilczek, Phys. Rev. D {\bf 8}, 3633 (1973).
\medskip

\bibitem{e-action}
N. K. Nielsen, Nucl. Phys. {\bf B101}, 173 (1975);\\
R. Fukuda and T. Kugo, Phys. Rev. D {\bf 13}, 3469 (1976);\\
C. Contreras and L. Vergara, hep-th/9610109.
\medskip

\bibitem{rg}
J. Kuti, L. Lin, and Y. Shen, Phys. Rev. Lett. {\bf 61}, 678 (1988).
\end{references}
\end{document}